\documentclass{PoS}
\usepackage{amsmath}
\usepackage{multirow}

\newlength{\bibitemsep}
\newlength{\bibparskip}
\let\oldthebibliography\thebibliography
\renewcommand\thebibliography[1]{%
  \oldthebibliography{#1}%
  \setlength{\parskip}{\bibitemsep}%
  \setlength{\itemsep}{\bibparskip}%
}

\title{Nucleon spin structure from lattice QCD}

\ShortTitle{Nucleon spin structure}

\author{Constantia Alexandrou\\
  Department of Physics, University of Cyprus, POB 20537, 1678 Nicosia, Cyprus,\\
  The Cyprus Institute, 20 Kavafi Str., Nicosia 2121, Cyprus\\
  E-mail: \email{alexand@ucy.ac.cy}}

\author{Martha Constantinou\\
       Department of Physics, Temple University, Philadelphia, PA 19122 - 1801, USA\\
       E-mail: \email{marthac@temple.edu}}

\author{\speaker{Kyriakos Hadjiyiannakou}\\
        The Cyprus Institute, 20 Kavafi Str., Nicosia 2121, Cyprus\\
        E-mail: \email{k.hadjiyiannakou@cyi.ac.cy}}

\author{Karl Jansen\\
       NIC, DESY, Platanenallee 6, D-15738 Zeuthen, Germany\\
       E-mail: \email{karl.jansen@desy.de}}

\author{Christos Kallidonis\\
  Department of Physics and Astronomy, Stony Brook University, Stony Brook, NY 11794, USA\\
       E-mail: \email{christos.kallidonis@stonybrook.edu}}

\author{Giannis Koutsou\\
  The Cyprus Institute, 20 Kavafi Str., Nicosia 2121, Cyprus\\
       E-mail: \email{g.koutsou@cyi.ac.cy}}

\author{Alejandro Vaquero Avil\'es-Casco\\
        Department of Physics and Astronomy, University of Utah, Salt Lake City, UT 84112, USA\\
        E-mail: \email{alexvaq@physics.utah.edu}}

\abstract{In this lattice QCD study we evaluate the nucleon spin decomposition to quarks and gluons contributions. We employ one gauge ensemble of maximally twisted mass fermions with two degenerate light quarks tuned to approximately reproduce the physical pion mass. We find that both spin sum and momentum sum are satisfied within the current statistical and systematic accuracy.}

\FullConference{XXVI International Workshop on Deep-Inelastic Scattering and Related Subjects (DIS2018)\\
		16-20 April 2018\\
		Kobe, Japan}

\begin{document}

\section{Introduction}
From 1987, since the European Muon Collaboration found that a small fraction of the nucleon spin is carried by quarks, the spin puzzle was triggered, which attracts an intense theoretical and experimental attention. Recent experiments find that the quark spin \cite{Ashman:1987hv} contributes to about 30\% of the nucleon spin, while there are indications of a significant gluon contribution \cite{Nakagawa:2017ljf}. Lattice QCD can measure directly the nucleon spin decomposition and therefore contribute significantly in our understanding about the nucleon structure. The valence quarks contribution has been measured in high accuracy on the lattice but a complete decomposition of the nucleon spin should take into account sea quarks and gluons contributions. The last two contributions are disconnected, showing a noisy behavior, therefore special techniques should be employed in combination with very large statistics, resulting in an increased computational cost. Recent algorithmic developments in combination with the increased computational power provided by GPUs enabled the accurate computation of all diagrams involved, allowing us to  provide a complete picture of the nucleon spin, as well as the momentum decomposition. For other studies see Refs \cite{Yang:2016plb,Yang:2018bft}.

\section{Lattice Extraction}
\subsection{Nucleon matrix elements}
One gauge invariant way to decompose the nucleon spin was introduced by X. Ji \cite{Ji:1996ek}, namely
\begin{equation}
J_N = \sum_q J_q + J_g = \sum_q \left( \frac{1}{2} \Delta \Sigma_q + L_q \right) +J_g.
\end{equation}
The intrinsic quark spin is $\frac{1}{2} \Delta \Sigma_q$ and $L_q$ is the quark orbital angular momentum. The total contribution of the gluons to the nucleon is $J_g$ where $J_g$ cannot decompose further, in contrast to the Jaffe-Manohar decomposition \cite{Jaffe:1989jz}. The intrinsic quark spin can be computed from the first Mellin moment of the polarized PDF and is the nucleon matrix element of the axial-vector operator at zero momentum transfer, $g_A^q \equiv \langle N(p,s') \vert A_\mu^q \vert N(p,s) \rangle = \bar{u}_N(p,s') \gamma_\mu \gamma_5 u_N(p,s)$, where $u_N(p,s)$ is the nucleon spinor with momentum and spin $p,s$. The total quark contribution $J_q$, can be extracted from the second Mellin moments of the unpolarized PDF or the Generalized Form Factors (GFFs) at zero momentum transfer, namely  $J_q = \frac{1}{2} \left[ A_{20}^q(0) + B_{20}^q(0) \right]$.
The GFFs can be extracted from the nucleon matrix element of the  vector one-derivative operator $\mathcal{O}^{\mu\nu}_V = \bar{q} \gamma^{\{ \mu } \overleftrightarrow{D}^{\nu\}} q $, namely $\langle N(p',s') \vert \mathcal{O}^{\mu\nu}_V \vert N(p,s) \rangle = \bar{u}_N(p',s') \Lambda^{\mu\nu}(Q^2) u_N(p,s)$
where $\Lambda^{\mu\nu}(Q^2) = A^q_{20}(Q^2) \gamma^{\{ \mu}P^{\nu\}} +B^q_{20}(Q^2) \frac{\sigma^{\{ \mu\alpha}q_{\alpha}P^{\nu\}}}{2m}+C^q_{20}(Q^2)\, \frac{1}{m}Q^{\{\mu}Q^{\nu\}}$ with $Q^2=(p'-p)^2$ is the momentum transfer square and $P=(p'+p)/2$. The $A_{20}(0)$ is directly accessible from the lattice data while $B_{20}(0)$ needs to be extrapolated at zero momentum transfer from finite $Q^2$ values.

In order to compute the $J_g$ term we construct the gluon operator \cite{Alexandrou:2016ekb}, $ \mathcal{O}^{\mu\nu}_g = 2 {\rm Tr} \left[ G_{\mu \sigma} G_{\nu \sigma} \right]$ where $G_{\mu\nu}$ is the field strength tensor. We use the scalar operator $\mathcal{O}_B = \mathcal{O}_{44} - \frac{1}{3} O_{jj}$ and from the matrix element $\langle N \vert \mathcal{O}_B \vert N \rangle = - 2 m_N \langle x \rangle_g$ one can extract $A_{20}^g(0) = \langle x \rangle_g$. To compute $J_g= \frac{1}{2} \left[A_{20}^g(0) + B_{20}^g(0) \right]$, evaluation of $B_{20}^g(0)$ is also needed. Assuming that spin and momentum sums are satisfied one can relate $B_{20}$ of gluons to quarks, such as $\sum_q B_{20}^q(0) = - B_{20}^g(0).$

\subsection{Simulation details}
In order to compute the spin and momentum decomposition of the nucleon we use one gauge ensemble of two mass degenerate twisted mass fermions with a clover term with mass tuned in order to approximately reproduce the physical pion mass. The lattice volume is $48^3 \times 96$ with lattice spacing $\alpha=0.0938(3)$ fm \cite{Abdel-Rehim:2015pwa}. For the strange and charm quarks we use Osterwalder-Seiler fermions where the strange and charm quark masses are tuned to reproduce the physical $\Omega^-$ and $\Lambda_c^+$ masses correspondingly \cite{Alexandrou:2017xwd}.

\subsection{Evaluation of the correlators}
The three-point functions $G_{\Gamma}^{\rm 3pt} (t_{ins},t_s)$ involved in the current calculation receive contribution from two classes of diagrams. When the insertion operator couples to a valence quark the diagram is called connected and when it couples to a sea quark is called quark-disconnected diagram. Note that strange quark and gluon contributions are purely disconnected. Typically, the dominant contribution comes from the connected diagram which can be computed using standard techniques such as sequential inversions through the sink. Computing disconnected diagrams is a challenge because on one hand the evaluation of the quark loop is computationally very expensive and on the other hand  disconnected quantities are  very noisy needing high statistics.

Simulations directly at the physical point are expensive due to the critical slow down of the inversion algorithm. In order to overcome the critical slow down we employ deflation of the lowest 500 modes of the twisted mass square operator to obtain more than 20x speedup for the light quark inversions. To estimate the quark loop we employ stochastic techniques and more specifically the one-end trick which yields an increased signal-to-noise ratio \cite{McNeile:2006bz}. One-end trick provides the quark loop for all the time-slices, therefore one can create the three-point function for any combination of $t_s,t_{ins}$. No dilution in the stochastic sources is employed. For the strange and charm quark inversions we employ the Truncated Solver Method (TSM) \cite{Bali:2009hu}. TSM allows for large number of low precision inversions corrected with a small number of high precision inversions to speedup the computation. For the computation of the gluon loops we apply stout smearing in the links entering the gluon operator to reduce UV fluctuations. The two-point functions produced for the quark-disconnected diagrams can also be used for the computation of the gluon momentum fraction.

\subsection{Isolation of the ground state}
The interpolating field with the quantum numbers of the nucleon creates also excited states which should be discarded in order to extract the matrix element of the ground state. In lattice QCD one creates a ratio of three- to two-point function to extract the matrix element,
\begin{equation}
  R_\mu(\Gamma_\nu,t_s,t_{\rm ins})  \xrightarrow[t_s-t_{\rm ins} \gg 1]{t_{\rm ins} \gg 1 \;\;}   \;\; \Pi_\mu(\Gamma_\nu)
  \label{Eq:Ratio}
\end{equation}
 but only for large enough time separations, the matrix element of the ground state is isolated. We employ three methods to assess the size of excited states and extract the ground state contribution.
\begin{enumerate}

\item \underline{Plateau:} Identifying a time-independent window in this ratio and extracting the desired matrix element by fitting to a constant is referred to as the plateau method. We seek convergence of the extracted value as we increase $t_s$.

\item \underline{Summation:} Summing over the insertion time $t_{ins}$ of the ratio in Eq. (\ref{Eq:Ratio}) we obtain
  \begin{equation}
    R^{\rm{sum}}_\mu(\Gamma_\nu,\vec{p}',\vec{p},t_s) \equiv \sum_{t_{\rm ins}=a}^{t_s-a} R_\mu(\Gamma_\nu,\vec{p}',\vec{p},t_s,t_{\rm ins})=C + t_s {\cal{M}} + \mathcal{O}(e^{-\Delta t_s}) + \cdots.
    \label{Eq:Summ}
  \end{equation}
  The constant $C$ is independent of $t_s$ and $\Delta$ is the energy gap between the first excited state and the ground state, while the matrix element of interest $\mathcal{M}$ is extracted from a linear fit to Eq. (\ref{Eq:Summ}) with fit parameters $C$ and $\mathcal{M}$.

\item \underline{Two-state fit:} Including the first excited state one has to fit the exponential terms to extract the masses and the amplitudes giving seven parameters in total. We perform a simultaneous fit to the three- and two-point functions to extract the matrix element of the ground state.
\end{enumerate}

\subsection{Renormalization}
We renormalize our lattice results using the RI'-MOM scheme and we subtract lattice artifacts using lattice perturbation theory. For details see Ref.~\cite{Alexandrou:2017hac,Alexandrou:2015sea}. The renormalization of the gluon operator is carried out perturbatively as explained in Ref.~\cite{Alexandrou:2016ekb}. All the following results are given in $\overline{\rm MS}$-scheme at 2 GeV.

\section{Results}
\begin{figure}[h!] 
  \begin{minipage}[t]{0.5\linewidth} 
    \includegraphics[width=\linewidth]{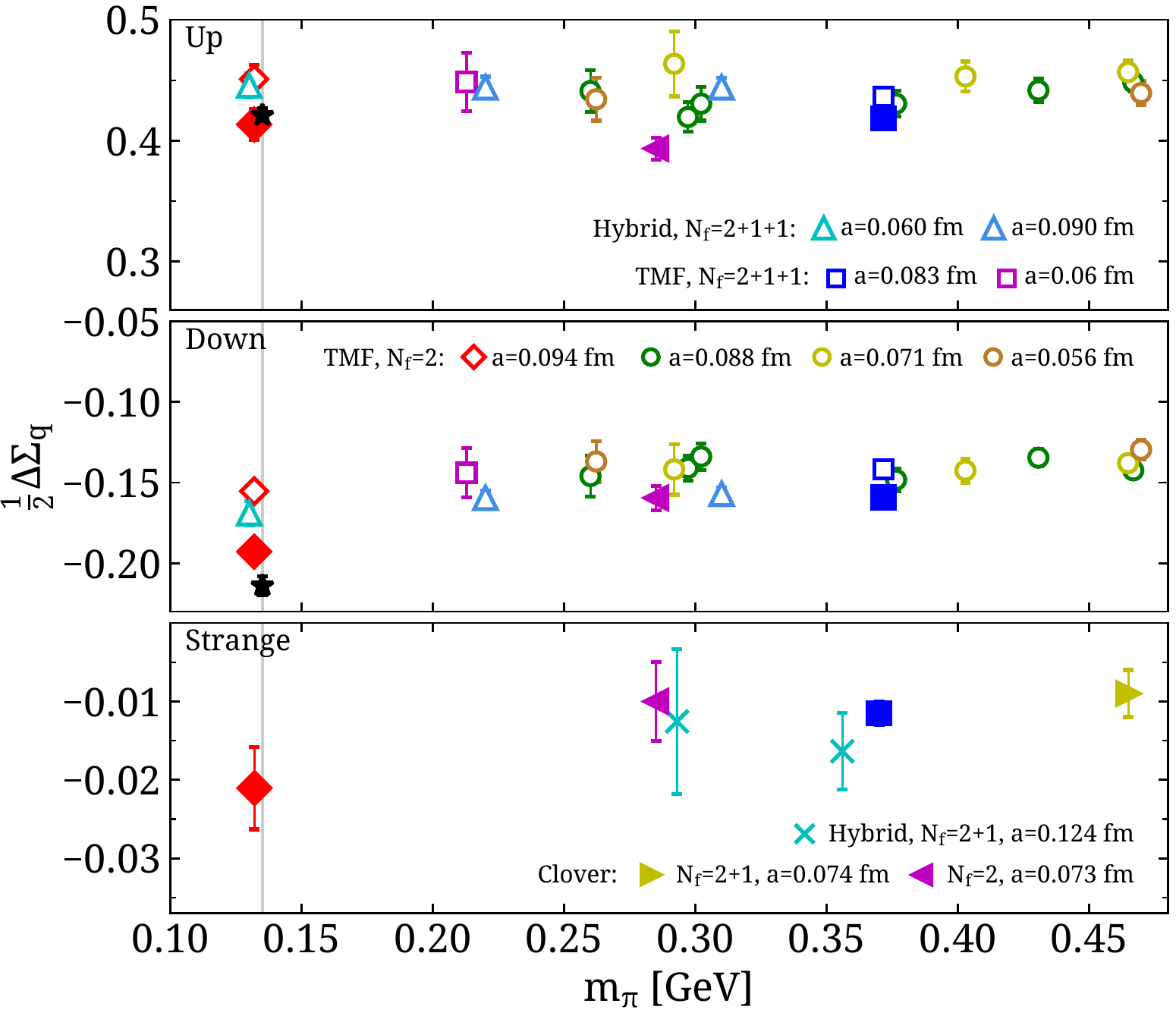}
  \end{minipage} 
  \hfill 
  \begin{minipage}[t]{0.5\linewidth} 
\vspace{-6.2cm}    \includegraphics[width=\linewidth]{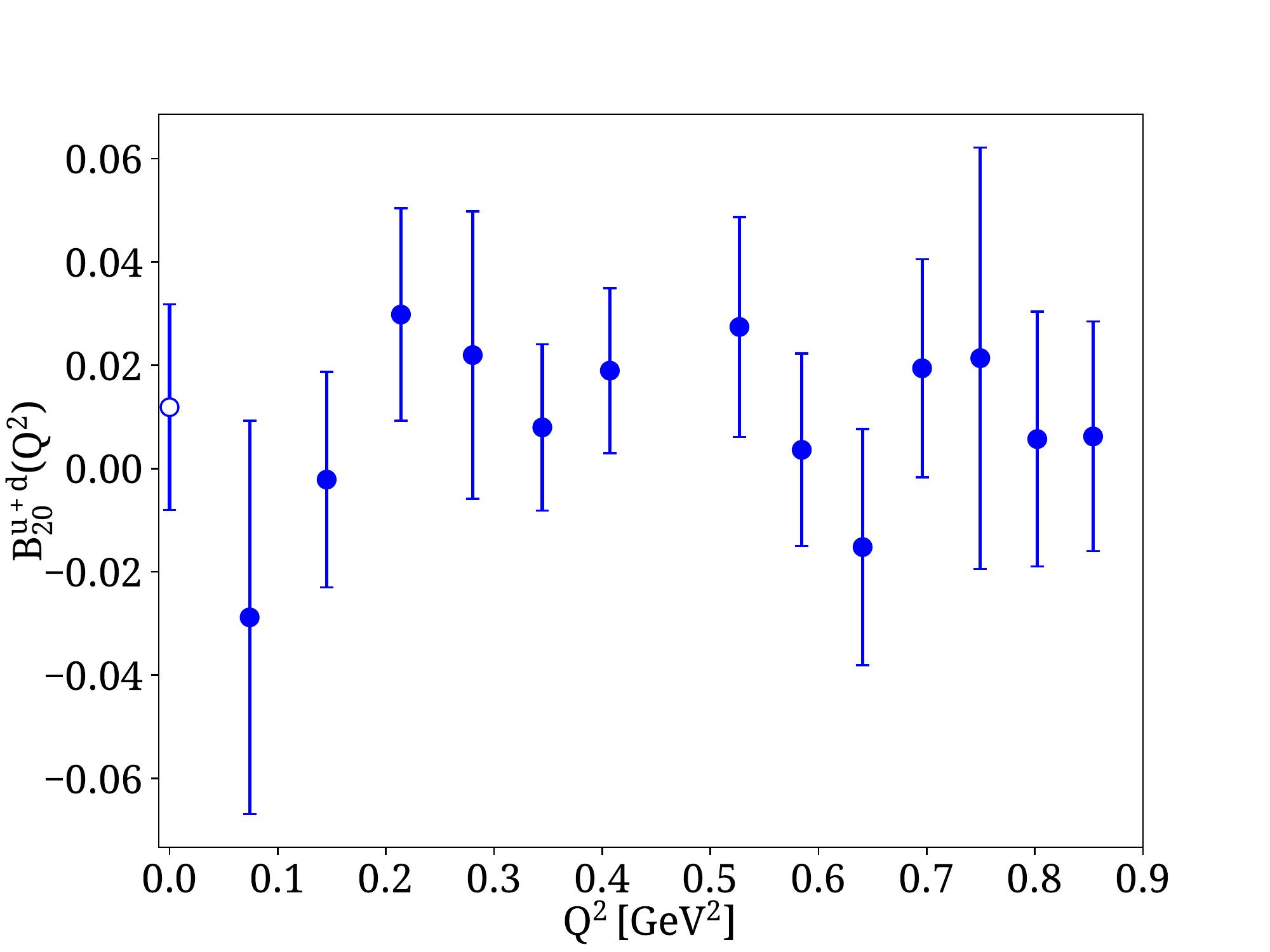}
  \end{minipage}
    \caption{\small Left: The up (upper), down (center) and strange (lower) quark intrinsic spin contributions to the nucleon spin versus the pion mass. Open symbols show results with only connected contributions while filled symbols results including disconnected. Red diamonds are the results of this work. Results from other studies are also presented. Right: The $B_{20}^{u+d}(Q^2)$ generalized form factor.} 
        \label{fig:DS_B20}
\end{figure}

In the left panel of Fig.~\ref{fig:DS_B20} we show our results for intrinsic quark spin contribution for the up, down and strange in comparison with other studies. It is worth mentioning that one has to include the contribution of the disconnected diagrams in order to find agreement with the experimental value of $\frac{1}{2} \Delta \Sigma_u$ and $\frac{1}{2} \Delta \Sigma_d$. Additionally, we highlight the fact that this quantity has very mild pion mass dependence. Since, we analyze only one ensemble, quenching effects, finite volume effects and finite lattice spacing effects cannot be accessed directly. Regarding cut-off effects our previous study with two lattice spacings, at the same pion mass, $m_\pi \sim 465$ MeV, and approximately the same spatial volume showed that cut-off effects are negligible. For the finite volume effects our previous study  at $m_\pi \sim 300$ MeV, for different $m_\pi L$ and same lattice spacing shows no finite volume effects as one can see from Fig.~\ref{fig:DS_B20}. For quenching strange quark effects we are currently analyzing a $N_f=2+1+1$ directly at the physical point and from preliminary results  we do not see quenching effects within the current statistical accuracy.

In the right panel of Fig.~\ref{fig:DS_B20} we present the isoscalar connected $B_{20}^{u+d}(Q^2)$ GFF. We find that in $Q^2$-range up to 1 GeV$^2$ its values are very small and compatible with zero. Using a naive linear extrapolation at zero momentum transfer we find that $B_{20}^{u+d}=0.012(20)$ small and compatible with zero. Additionally we find that disconnected contributions are much smaller and negligible. Therefore, we expect that $B_{20}^g$ should not contribute and we thus use that $J_g = \frac{1}{2} \langle x \rangle_g$.

\begin{figure}[ht!]
  \begin{minipage}[t]{0.45\linewidth} 
    \includegraphics[width=\linewidth]{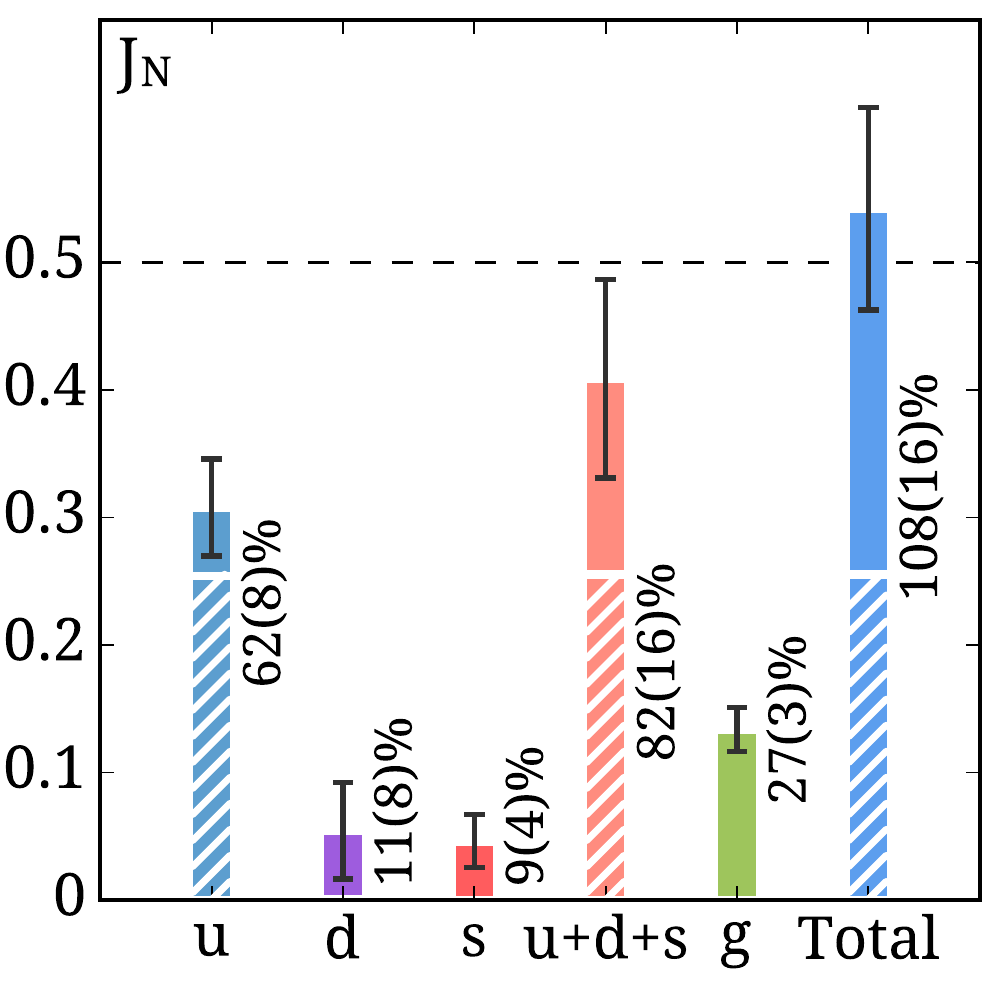}
  \end{minipage} 
  \hfill 
  \begin{minipage}[t]{0.45\linewidth} 
    \includegraphics[width=\linewidth]{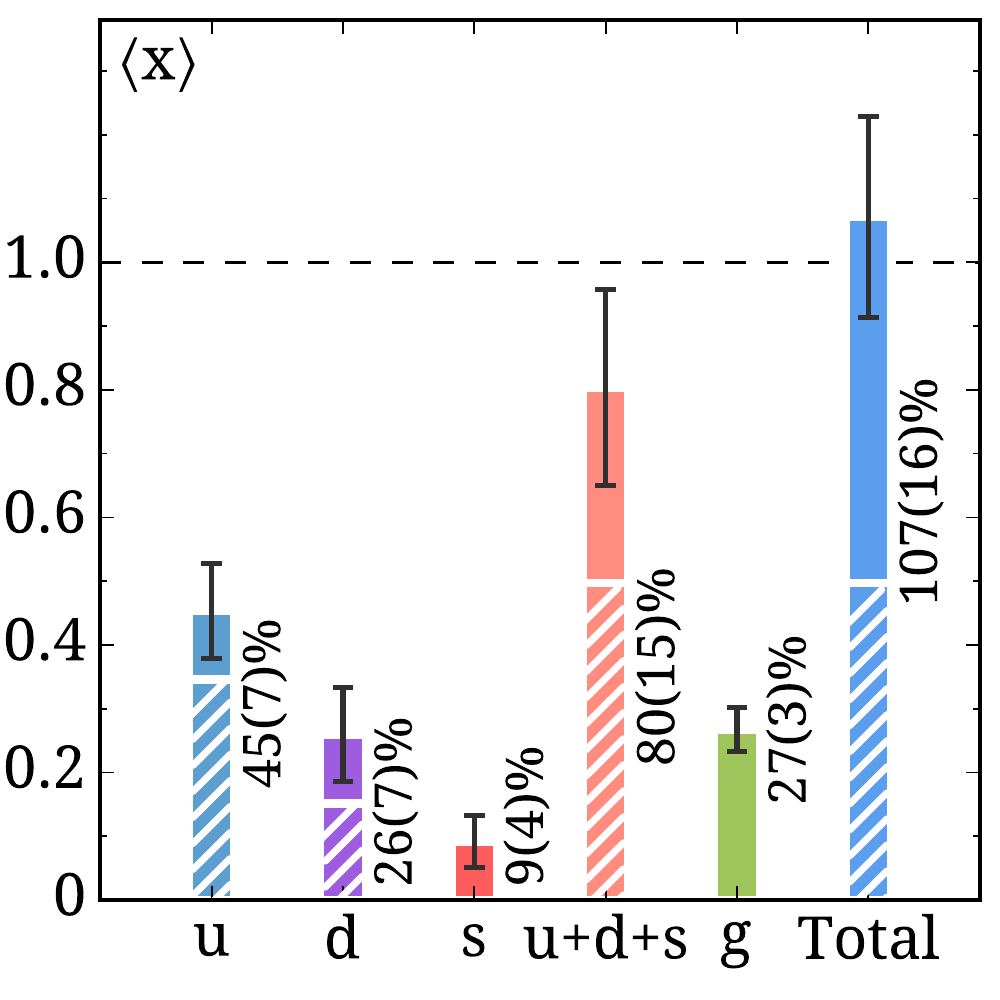}
  \end{minipage}  
   \vspace*{-0.4cm}
  \caption{ Left: Nucleon spin decomposition. Right: Nucleon momentum
    decomposition. The striped segments show valence quark
    contributions (connected) and the solid segments the sea quark and gluon
    contributions (disconnected). Results are given in $\overline{\rm MS}$-scheme at 2 GeV.}
  \label{fig:pies}
\end{figure}

In Fig.~\ref{fig:pies} we present our results for the complete nucleon spin and momentum decomposition \cite{Alexandrou:2016ekb}. We find that disconnected diagrams contribute significantly and that the spin and momentum sums are satisfied within the current statistical accuracy. All the results are summarize in Table \ref{table:quark spin}. 

\begin{table}
\begin{center}
  \caption{Our results for the intrinsic spin
    ($\frac{1}{2}\Delta\Sigma$), angular  ($L$) and total
    ($J$) momentum contributions to the nucleon spin and to the nucleon
    momentum $\langle x \rangle$, from quarks and gluons,
    where the first error is statistical and the second a systematic
    due to excited states.}
  \label{table:results}
  \begin{tabular}{rr@{.}lr@{.}lr@{.}lr@{.}l}
    \hline\hline
    & \multicolumn{2}{c}{$\frac{1}{2}\Delta\Sigma$} & \multicolumn{2}{c}{$J$} & \multicolumn{2}{c}{$L$}  & \multicolumn{2}{c}{$\langle x \rangle$} \\\hline
          u& 0&415(13)(2)        & 0&308(30)(24)& -0&107(32)(24)        & 0&453(57)(48)\\
          d&-0&193(8)(3)         & 0&054(29)(24)&  0&247(30)(24)        & 0&259(57)(47)\\
          s&-0&021(5)(1)         & 0&046(21)(0) &  0&067(21)(1)         & 0&092(41)(0) \\
          g&\multicolumn{2}{c}{-}& 0&133(11)(14)&  \multicolumn{2}{c}{-}& 0&267(22)(27)\\
       tot.& 0&201(17)(5)        & 0&541(62)(49)&  0&207(64)(45)        & 1&07(12)(10) \\\hline\hline
  \end{tabular}
  \label{table:quark spin}
  \end{center}
\end{table}

\section{Conclusion} \vspace{-0.5cm}
We have presented a complete study of the nucleon spin and momentum decomposition to quark and gluon contributions using one $N_f=2$ twisted mass ensemble directly at the physical point. Systematic effects assessed from ensembles with heavier pion masses, therefore we expect systematics smaller than the statistical uncertainty. For the intrinsic quark spin we find $\frac{1}{2} \Delta \Sigma_{u+d+s} = 0.201(17)(5)$ for the nucleon spin $J_N=0.541(62)(49)$ and for the momentum sum $\sum_q \langle x \rangle_q + \langle x \rangle_g = 1.07(12)(10)$.

\vspace{-0.5cm}
\bibliographystyle{JHEP}
\bibliography{refs}

\end{document}